# Stochastic Dynamics for Earthquake Ruptures

Tsung-Hsi Wu[1]*and Chien-Chih Chen[1,2]†

[1]Department of Earth Sciences, National Central University, Jhongli 32001, Taiwan

also at:

[2]Earthquake-Disaster & Risk Evaluation and Management Center, National Central University, Jhongli 32001, Taiwan

Corresponding author: Tsung-Hsi Wu (tsung.hsi@g.ncu.edu.tw)

**Abstract**

In this paper, we propose a stochastic dynamic model for earthquake rupture and suggest that the Langevin equation of frictions may be used for interpreting the slip distributions of rupture processes in earthquakes. The steady-state solution of the derived Langevin equation analytically attains the truncated exponential (TEX) distribution that is empirically characterized in many rupture models of earthquake events worldwide, as demonstrated by Thingbaijam and Mai (2016, https://doi.org/10.1785/0120150291). Our proposed stochastic dynamic faulting model for earthquake rupture intrinsically includes fluctuations and uncertainties in the heterogeneity of faulting planes as random variables. Specifically, we related the characteristic parameter $u_c$ in TEX functions to the ratio of diffusion and friction coefficients $D$ and $\gamma$ of the Langevin equation.

## Introduction

Modern earthquake rupture models that provide us the source slip motion are mainly obtained through inverting observed ground motion collected by near-field, regional-distance (2°–12°), and far-field seismograms. Earthquake rupture inversion is essentially an optimal process of solving a nonlinear problem by searching in parameter space exhaustively to minimize the residuals to the data (Meyers, 2011, pp. 85–86). Mathematically, it is a deterministic process of solving vector-matrix equations. Owing to the general underdetermined nature of the inversion problem, arbitrary parameter choices and many simplifications must be made with geodetic or ground-shaking observations as constraints in slips on fault planes, to render such an inverse problem tractable (Papageorgiou, 2003; Siriki et al., 2015). Therefore, several uncertainties are ignored while solving deterministic seismic inversion problems. Furthermore, conducting rigorous time-dependent stress analysis is difficult, slip functions are usually chosen intuitively. Accordingly, most rupture models are kinematic rather than dynamic.

In practice, stress in the rupture process is calculated using constitutive laws or derived from particle slip/velocity histories of a kinematic model. The vast number of degrees of freedom in a faulting system make it difficult to analyze the forces in an earthquake rupture process. Theoretically, we can predict the evolution of any macroscopic dynamical system deterministically with classical mechanics, but for systems with many degrees of freedom, differential equations become explicitly unsolvable (Eckmann, 1981). Moreover, the presence of many degrees of freedom in a dynamical system leads to deterministic chaos, which is practically indistinguishable from that produced by a stochastic process (Cecconi et al., 2005).

The study of stochastic process was initiated in 1905, when Einstein successfully interpreted the irregular movement of Brownian particles by solving a partial differential equation, that is, the Fokker–Planck equation (FPE), governing the time evolution of the probability density of Brownian particles. After 3 years, Langevin devised a stochastic differential equation (SDE) called the Langevin equation, which is the first successful dynamical model for Brownian motion. Both the Langevin equation and FPE describe the physics of continuous and memoryless Markov processes (Lemons and Gythiel, 1997; Renn, 2005). In its simplest form, the Langevin equation is the equation of motion for a system that experiences a particular type of random force in the environment.

Unlike the deterministic equation of motion, the Langevin equation can never predict a particle's trajectory. Particular solutions of Langevin equation have little meaning until a statistical analysis is performed. The capability of the Langevin equation in modeling the statistical behavior of Brownian particles as well as other open systems makes it useful for describing a variety of physical process, such as those of supercooled liquid, the human brain, finance, meteorology, earthquake faults, nuclear physics, and granular matter (Yulmetyev et al., 2009; Eslamizadeh and Razazzadeh, 2018).

For earthquake rupture processes, the interactions between the entities composing the system are too complex such that obtaining analytical molecular dynamics solutions is impossible, and only limited numerical solutions are available (Rundle et al., 2003). This situation is similar to the case of Brownian motion. Inspired by Brownian motion, this study constructs a stochastic dynamic model capable of generating the equivalent process for earthquake rupturing. On the basis of two candidate Langevin equations and their corresponding FPEs, we generated stochastic rupture processes numerically and derived the probability density function (PDF) of slips analytically. Both the analytical solution of FPE and the numerical solution of the Langevin equation coincide with the evidence for the truncated exponential (TEX) probability distribution of earthquake slip (Thingbaijam and Mai, 2016). The proposed Langevin equations, which omit many degrees of freedom and the complexity of the faulting system, describe the possible mechanisms of rupture processes in a simple and physically reasonable manner.

# Stochastic Dynamics for Heterogeneous Frictions

*Langevin equation of frictions*

To describe Brownian motion, Langevin applied Newton's second law to a representative Brownian particle. In his description, the process is assumed to be memoryless and nondifferentiable (Hänggi and Marchesoni, 2005), and the overall stochastic force can therefore be subdivided into a deterministic damping force and an independent random fluctuating force, as shown in Eq. (1).

$$\frac{dv}{dt} = -\gamma v + \Gamma(t) \qquad (1)$$

In Eq. (1), $v = dx/dt$ is the velocity of the particle; $m$ the mass; $\gamma$ the frictional coefficient per unit mass; $-\gamma v$ the damping force, representing a viscous friction environment; and $\Gamma(t)$ the random force applied on the particle per unit mass, a time-varying fluctuating force with an origin in the uncertainty of the environment. Similarly, the Langevin equation for a Brownian particle under Coulomb friction (Hayakawa, 2005) is written as

$$\frac{dv}{dt} = -\gamma \frac{v}{|v|} + \Gamma(t). \qquad (2)$$

The first term on the right hand side of Eq. (2) is the Coulomb dragging force caused by the background friction. Similar to ordinary differential equations, SDEs [Eqs. (1) and (2)] are used as the equations of motion describing the Brownian particle pushed by the incessantly fluctuating force $\Gamma(t)$ and subjected to a viscous damping force [Eq. (1)] or a constant Coulomb frictional force [Eq. (2)].

For a Brownian particle under a viscous friction and an additional external force $g(x)$, the Langevin equation can be written as follows (Risken, 1989):

$$\frac{d^2x}{dt^2} = -\gamma \frac{dx}{dt} + \frac{g(x)}{m} + \Gamma(t) \tag{3}$$

Assuming $\gamma$ is sufficiently large, we applied an overdamped approximation scheme for Eq. (3), and it subsequently becomes

$$\frac{dx}{dt} = \frac{g(x)}{m\gamma} + \frac{\Gamma(t)}{\gamma}. \tag{4}$$

Eq. (4) describes the Brownian motion in an overdamping viscous friction regime. If we define $g(x) \equiv -kmx/|x|$, that is, a constant external force pulling particles toward a certain equilibrium state, we have a Langevin equation similar to Eq. (2):

$$\frac{dx}{dt} = \frac{-kx}{\gamma |x|} + \frac{\Gamma(t)}{\gamma}. \tag{5}$$

Note that Eq. (5) is a Langevin equation of displacement as a function of time, whereas Eq. (2) is a Langevin equation of velocity as a function of time, for a representative Brownian particle. Eqs. (2) and (5) are equivalent from mathematical point of view, but their physical scheme and meaning are different.

In practice, the random fluctuating force $\Gamma(t)$ is conventionally assumed to be white noise defined by the time derivative of the Wiener process $W(t)$, and it satisfies the following equation (Kubo, 1966; Jazwinski, 1970, p. 85):

$$<\Gamma(t)\Gamma(s)> = 2D\delta(t - s) \tag{6}$$

where $2D$ is the spectral density of $\Gamma(t)$. Although the background fluctuation always has some memory effect in reality, it is still usually approximated by a delta function (no memory) in many fields of science and engineering as well as in this study to enable analytical calculation.

Considering the rupture process in real earthquakes, the medium where the rupture front propagates is usually highly heterogeneous, where asperities, cracks, and fractures may be regarded as randomly distributed. Thus, the stress level at the rupture front fluctuates widely, causing a highly heterogeneous slip distribution over the fault in most observations or source models. Moreover, in contrast to the perfectly elastic medium (in which the rupture process does not cease until it reaches one of the boundaries), the plastic-elastic crust provides a dissipative force that impedes the particle motion. Langevin equations [Eqs. (2) and (5)] are proposed to model the particle motion at the rupture front of a realistic rupture process in a stochastic manner, in which the fluctuation of stress level at the rupture front is described as the result of a random force $\Gamma(t)$, and the dissipative force is controlled by a frictional coefficient γ. The propagation of the rupture is regarded as a series of momentum transfers between adjacent particles.

In the later section, the Langevin equation is numerically solved, generating a time series $\{(t_i, x_i)\}$ as a realization of the stochastic process. We assume that the rupture propagates at constant velocity $V_R$, and hence, the generated time series indicates that the rupture slip has a value $x_i$ occurring at distance $r_i \equiv t_i V_R$ from the hypocenter point. The random fluctuation $\Gamma(t)$ is defined as Gaussian white noise with constant strength $D$ and independent of time (to allow analytical calculation). As a consequence of these assumptions, the model we built is only valid for the rupture process in the stationary state, which means it may fail when the rupture is about to end or rest. The mechanism of triggering a rupture process or stopping a process is out of the scope of this study.

*FPE and the Smoluchowski equation*

For SDEs, we can only determine the expectation value of the random variable as a function of time by using the Ito stochastic integral or obtaining a particular solution (also called a trajectory, a sample path, or a realization) through numerical methods. In contrast to an ordinary differential equation, each sample path of an SDE varies considerably even under the same initial conditions. Physically, the trajectory of a Brownian particle governed by a Langevin equation varies each time we observe the particle. To obtain the probability density as a function of time and space of the Brownian particle, we must repeatedly solve the SDE numerically and perform statistical calculation, and then we obtain only an approximation of the PDF of the random variable. Such a process is limited; thus, the FPE is required.

The FPE describes the evolution of the distribution function of fluctuating macroscopic variables. The well-known diffusion equation that describes the collective behavior of an assembly of free Brownian particles is a simple example of the FPE (Kampen, 1981, p. 209; Coffey et al., 1996, p. 45). The FPE is an approximation from the master equation, which is a partial differential equation of the transition probability of each state in a given Markov process (Kampen, 1981). For a stochastic process defined by a Langevin equation

$$\frac{dX(t)}{dt} = A(t, X(t)) + \Gamma(t), \qquad (7)$$

we have the corresponding FPE (Kampen, 1981, p. 244),

$$\frac{\partial P(x,t)}{\partial t} = -\frac{\partial}{\partial x}\Big(A(t, X(t))P(x,t)\Big) + D\frac{\partial^2}{\partial x^2}\big(P(x,t)\big). \qquad (8)$$

The arbitrary linear mapping function $A(t, X(t))$ is called the drift term, and the constant $D$ that is obtained from Eq. (6) is called the diffusional coefficient. By solving the FPE, the PDF $P(x, t)$ can be obtained, which reads as the probability density of observing the stochastic process $X(t, \omega)$ having the value $x$ at time $t$. The form of the FPE in Eq. (8) is also called the Smoluchowski equation, which is a special form of the generalized FPE, developed by Smoluchowski in 1906 independently from Einstein for finding the PDF of the physical quantities of a Brownian particle in a viscous environment (Coffey et al., 1996, p. 13; 54).

*Equivalence of the Langevin equation and FPE*

The Langevin equation and FPE are mathematically equivalent (Kampen, 1981, p. 209). They both describe Markovian random walks but from different approaches. In a numerical simulation, the Langevin equation provides a particular solution of the stochastic process, whereas the FPE solution, which is a PDF, is the statistical average of infinite individual solutions of the corresponding Langevin equations. In practice, only a few FPEs can be solved analytically, whereas obtaining the time evolution of the PDF by solving a Langevin equation numerically is time consuming (Zorzano et al., 1999). In this paper, we use Langevin equations for simulating the rupture process and for their physical interpretation. FPE is used for obtaining the PDFs of rupture slips in the stationary state.

## Essential Properties of the Markovian Stochastic Process

*Equivalence of time average and ensemble average*

According to the ergodic hypothesis of Boltzmann, if the system reaches ergodicity, the time average is equal to the ensemble average (i.e., statistical expectations) (Plato, 1991; Coffey et al., 1996), as shown in Eq. (9).

$$\mathbb{E}[X(t)X(t+\tau)] = \overline{X(t)X(t+\tau)} = \lim_{T \to \infty} \frac{1}{T} \int_0^\infty X(t)X(t+\tau)dt \qquad (9)$$

The ergodic hypothesis of Boltzmann was later proved to be not strictly true (Gardiner, 1985). However, for a stationary process in ergodicity, all time-dependent averages are functions only of time differences; the ensemble average and time average always give the same result (Gardiner, 1985; Coffey et al., 1996). That is, for a stationary process, the ergodic hypothesis of Boltzmann holds true. The equivalence of the time average and ensemble average allows us to testify the analytical stationary solution with the empirical observation later.

*Ruptures as Markov processes*

Consider a stochastic process $X = \{X(t), t \geq 0\}$, with its sample path denoted by $\{X(t_1) = x_1, X(t_2) = x_2, \ldots, X(t_N) = x_N\}$. The process is called a Markov process if the conditional probability density of $x$ at $t_n$ is uniquely determined according to $x_{n-1}$ at $t_{n-1}$ and is not affected by any knowledge of an earlier time (Van Kampen, 1992). In brief, the transition probability only depends on the present state, that is

$$P_{1|n-1}(x_n|x_1, x_2, \ldots, x_{n-1}) = P_{1|1}(x_n|x_{n-1}). \tag{10}$$

A stochastic process obeying the proposed Langevin equation [Eq. (2) or (5)] is a time-invariant Markovian process, as $Pr\{X_{n+1} = b | X_n = a\} = Pr\{X_2 = b | X_1 = a\}$ is always satisfied for any $a, b$, and $n$. The invariant conditional probability of transition over any two adjacent states means that one sample path of the Langevin equation can also be regarded as a collection/combination of individually generated segments each with zero start and end values. Due to the memoryless property of the Markov process, the order of segments can be arbitrarily rearranged without losing equivalence if the value at the end of one segment meets the initial value of its succeeding one.

In this study, the earthquake rupture processes are assumed to be governed by the Langevin equation [Eq. (4) or (5)] in which negative slips will occur as frequently as positive slips in general. Conversely, observations have shown that ruptures usually favor a certain direction in an earthquake event. This problem vanishes as we add two simple assumptions. First, the idea of an earthquake event is assumed to be a collection of subrupture process (subevents). Second, we assume that each subrupture process starts with zero slip and terminates immediately as it reaches zero again. Because the stochastic rupture process governed by the proposed Langevin equation is Markovian, a sample path for the rupture process can be constructed by individually generated sample paths, each having zero start and zero end. With an additional negligibly small constant bias/external force in the Langevin equation, the overall rupture slips will always favor a certain direction.

## Data Processing for the Rupture Model

*Numerical simulation of the stochastic rupture process*

In this study, particular solutions of the corresponding SDE to the Langevin equations are solved in Euler scheme following the instruction manual of Cyganowski et al. (2001). Two parameters (i.e., diffusional coefficient $D$ and frictional coefficient $\gamma$) in both Langevin equations [Eq. (2) and (5)] control the process evolution. Duration $T$ gives the total number of steps ($N-1$) of one sample path through $T/dt$ (rounded to the nearest integer). Time step is defined by $dt = 10^{-5}$ arbitrary unit (arb. unit) of simulation time, and $N$ is the total number of states (data points) in a sample path. The unit/scaling of simulation time depends on how we define the unit of coefficient $D$ and $\gamma$.

To demonstrate simulated results regarding the widely observed TEX character of earthquake slip distributions (Thingbaijam and Mai, 2016), the distribution of the realization values of a sample path is cumulatively calculated, subtracted from 1, and then fitted with the complementary cumulative distribution function (CCDF) of the TEX and exponential (EXP) function. The truncated distribution function is a conditional distribution obtained by restricting the domain of the original function. It preserves the main features of the original function and avoids extreme values being involved unrealistically. Fitting data to the function in CDF form is for avoiding the arbitrary choice of data binning that may greatly affect the fitting result. CCDF is simply a complement to CDF, presenting the distribution in a similar manner as the cumulative frequency-magnitude distribution (Cosentino et al., 1977, as cited in Thingbaijam and Mai, 2016).

Consider an EXP distribution $P_0 exp(-u/u_c)$ being truncated within the range [0, $u_{max}$] and renormalized, where $u_{max}$ is the maximum value in the data, and we have the TEX distribution:

$$f(u) = \frac{1}{u_c} \frac{exp(-\frac{1}{u_c}|u|)}{1 - exp(-\frac{1}{u_c}u_{max})}. \tag{11}$$

Correspondingly, the CCDF of TEX will be

$$1 - F(u) = \frac{exp\left(-\frac{1}{u_c}|u|\right) - exp\left(-\frac{1}{u_c}u_{max}\right)}{1 - exp\left(-\frac{1}{u_c}u_{max}\right)}, \tag{12}$$

where $F(u)$ is the cumulative distribution function of the TEX density function $f(u)$.

In this study, the TEX distribution of a finite-source rupture model is used for comparison. The $u_c$ obtained from the best-fitting TEX function to the slip distribution of this reference model provides the $D/\gamma$ ratio that constrains the generation of sample paths. The relation $u_c \equiv D/\gamma$ results from the analytical solution of FPE, the details of which are presented in the *Analytical solution of the corresponding FPE* section.

Because the stochastic rupture processes governed by the proposed Langevin equations are time-invariant Markovian, we simply removed the negative sign of the sample path (which is a series of realization values) before data fitting to EXP and TEX functions. A sign-free time series of duration $T$ generated by the SDE is equivalent to a set of segments individually generated according to the same Langevin equation with a very small positive bias. The summation of the time interval of each segment is equal to $T$.

*Particular solution of the Langevin equation*

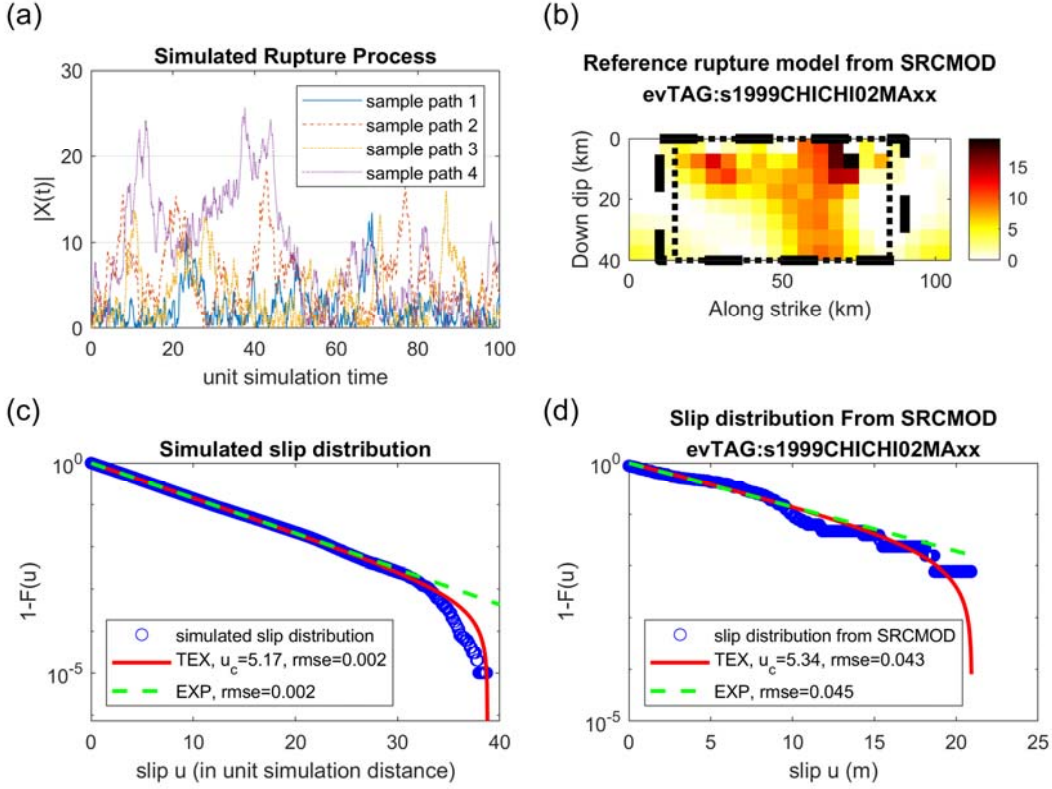

Figure 1. CCDF of 100 particular solutions of the stochastic earthquake rupturing, with the event (s1999CHICHI02Maxx) from SRCMOD as a reference. The red solid lines (TEX) and green dashed lines (EXP) are the best-fitting functions to slip distributions (blue circles). (a) Four of the total 100 sample paths of the process X(t). (b) The slip distribution of the reference rupture model. (c) The slip distribution of the simulated stochastic rupture process X(t). (d) The slip distribution (in CCDF) of the reference rupture model. The color version of this figure is available only in the electronic edition.

In the *Langevin equation of frictions* section, we proposed the mathematically equivalent Langevin equations (2) and (5) for the earthquake rupture process, and thus the corresponding solutions should be the same despite the difference in coefficients. Herein, we provide the equivalent SDE for Eq. (2), with $\Gamma(t) \equiv \sqrt{2D}\, dW(t)/dt$ satisfying Eq. (6):

$$dX(t) = -\gamma \frac{X(t)}{|X(t)|} dt + \sqrt{2D}\, dW. \quad (13)$$

For Eq. (5), the SDE has the same form, only with $\gamma$ and $D$ in Eq. (10) replaced by $\gamma/k$ and $D/\gamma^2$ respectively. Particular solutions in this study were numerically solved in Euler scheme. Figure 1 shows the TEX fitting result of a total of 100 sample paths generated according to Eq. (13) under the same initial conditions and parameters. Duration $T$ of each sample path was set to 100 (arb. unit). The initial condition was set to be zero, with a frictional coefficient $\gamma = 1$ for simplicity and a diffusional coefficient $D = 5.34$ according to the $u_c$ of the reference rupture model from the Finite-Source Rupture Model Database (SRCMOD, an online database of earthquake source models, see the Data and Resources section). The reference model is the source inversion of the 1999 Chi-Chi earthquake in Taiwan (Ma et al., 2001), a typical example of a large earthquake in which the slip distribution follows the TEX law. The simulation result shows that the distribution of a random variable $X(t)$ also follows the TEX law.

Regarding real rupture models resulting from source inversions, microscopic details are practically unavailable because high-frequency seismic waves decay very fast, resulting in spatial resolutions of hundreds of meters to kilometers. On a predefined two-dimensional (2D) finite fault plane, the slip value can be regarded as the average contribution of the subrupture processes that occurred in a cell of the grid size. Conversely, solving the Langevin equation for rupture dynamics numerically generated 100 one-dimensional sample paths, each containing tens to hundreds of subsegments. The cumulative slip calculation of these sample paths avoids the problem of dimensionality and roughly reproduces the slip distribution of a rupture event in a sense of averaging the contributions of different realizations. However, the issue of scaling is not discussed in this study.

*Analytical solution of the corresponding FPE*

Solving the corresponding FPE can yield the PDF of $X(t)$ of a stochastic process. The corresponding FPE to the Langevin Eq. (2) is

$$\frac{\partial P(v,t)}{\partial t} = \frac{\partial}{\partial v}\left[\gamma \frac{v}{|v|} P(v,t)\right] + \frac{\partial^2}{\partial v^2}[DP(v,t)]. \tag{14}$$

Similarly, the corresponding FPE to the Langevin Eq. (5) is

$$\frac{\partial P(x,t)}{\partial t} = \frac{\partial}{\partial x}\left[\frac{k}{\gamma} \frac{x}{|x|} P(x,t)\right] + \frac{\partial^2}{\partial x^2}\left[\frac{D}{\gamma^2} P(x,t)\right]. \tag{15}$$

Considering the stochastic process in a stationary state, where the PDF is stable and time invariant, Eqs. (14) and (15) can be approximated to Eqs. (16) and (17), respectively (Kawarada and Hayakawa, 2004; Hayakawa, 2005).

$$-\frac{v}{|v|} P_{st}(v) = \frac{\partial}{\partial v}\left[\frac{D}{\gamma} P_{st}(v)\right]. \tag{16}$$

$$-\frac{x}{|x|} P_{st}(x) = \frac{\partial}{\partial x}\left[\frac{D}{k\gamma} P_{st}(x)\right]. \tag{17}$$

where $P_{st}$ is the PDF of particle velocity $v$ or dislocation $x$ for the process in stationary state. The solutions of Eqs. (16) and (17) are therefore

$$P_{st}(v) = P_0 \exp\left(-\frac{\gamma}{D}|v|\right). \tag{18}$$

$$P_{st}(x) = P_0' \exp\left(-\frac{k\gamma}{D}|x|\right). \tag{19}$$

The aforementioned EXP distributions [Eqs. (18) and (19)] are the analytical solutions of FPEs [Eqs. (14) and (15)] under stationary approximation, representing the theoretical distribution of the slip velocity or displacement at the front of the stochastic rupture process in the stationary state, where $P_0$ and $P_0'$ are arbitrary constants.

Furthermore, as we truncate and calculate the CCDF of $P_{st}$ in Eqs. (18) and (19), we have

$$1 - F(v) = \frac{exp\left(-\frac{\gamma}{D}|v|\right) - exp\left(-\frac{\gamma}{D}v_{max}\right)}{1 - exp\left(-\frac{\gamma}{D}v_{max}\right)} \qquad (20)$$

and

$$1 - F(x) = \frac{exp\left(-\frac{k\gamma}{D}|x|\right) - exp\left(-\frac{k\gamma}{D}x_{max}\right)}{1 - exp\left(-\frac{k\gamma}{D}x_{max}\right)}, \qquad (21)$$

respectively, in which the constants $P_0$ or $P_0'$ are naturally eliminated during the truncation.

The solution of the FPE is an ensemble-averaged PDF. That is, $P(u,t)$ as a solution of the FPE represents the PDF of a certain quantity having value $u$ at time $t$. For observations in seismology, however, we can never obtain enough samples for meaningful and reliable ensemble-averaged statistics because a sufficiently large number of repeated earthquakes triggered under the same or even similar conditions do not exist. Available empirical data such as slip distributions in SRCMOD are calculated in a time-averaged manner (i.e., the distribution is calculated according to values of one realization across time).

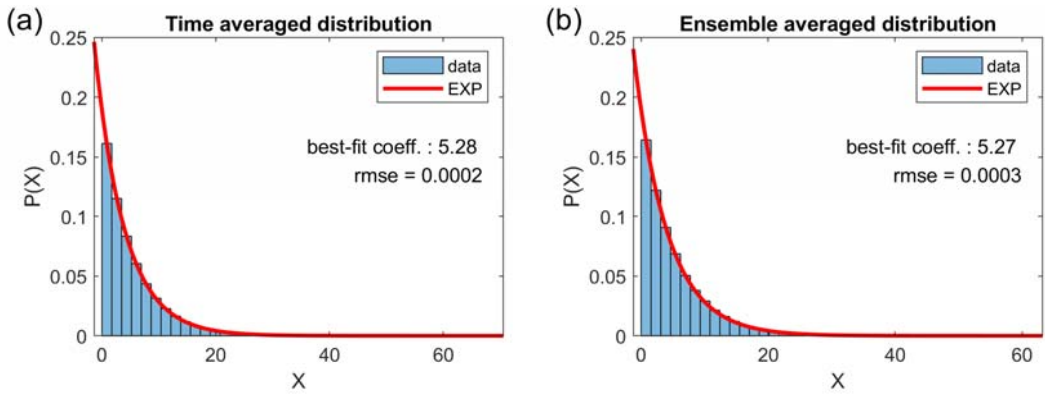

Figure 2. Histograms (probability density distributions) and the corresponding best-fitted EXP curve. (a) The distribution of the samples of a sample path with a duration of $3 \times 10^6$ (arb. unit). (b) The distribution of $X(t)$ at $t = 30$, samples from total $10^5$ individually generated sample paths.

According to the ergodic hypothesis of Boltzmann, the time-averaged PDF equals the ensemble-averaged one for the stochastic process in the stationary state (Gardiner, 1985; Coffey et al., 1996). Therefore, the stationary solution of the FPE can take the place of the time-averaged statistics of a stationary process. Figure 2 displays the best-fitted EXP curve [in terms of Eqs. (18) or (19)] to the distributions of (a) samples in a long sample path of total duration $3 \times 10^6$ (arb. unit), and (b) samples at $t = 30$ (arb. unit) of total $10^5$ realizations. The samples are generated using Eq. (13) with $D$ = 5.34 and $\gamma$ = 1. The result shows that the ensemble-averaged distribution is almost identical to the time-averaged one that approximates the FPE prediction well.

However, as in Eq. (9), the equality/equivalence of time- and ensemble-averaged distribution strictly holds for $T \to \infty$, and thus the EXP distribution as the solution of Eq. (18) or Eq. (19) is for processes of infinite duration. For statistical data from real observations, such as rupture slips in an earthquake, as well as numerical simulations with a finite duration, a finite maximum value always exists, and the distribution is therefore a truncated one. Hence, empirical results of SRCMOD or the numerical simulations in this study follow the TEX distribution rather than EXP. The smaller-than-expected TEX parameter $u_c$ of the simulation results (displayed in Figure 1 and 2, both with $u_c$ < 5.34) is due to the finite duration of the sample path.

In this section, we demonstrated that the stochastic rupture slips obtained through the Langevin equation follow the TEX distribution, which exhibits similar statistical behavior as the reference rupture model. Furthermore, we obtained the theoretical PDF of stochastic rupture slips by solving the corresponding FPE. The result shows that the PDF of rupture slips for an infinite stochastic process is originally EXP. With the existence of a finite maximum value, the empirical or simulated slip distribution is therefore TEX rather than EXP.

**Discussion**

Seismologists have long attempted to locate a specific function best describing the slip distribution along a fault plane during an earthquake rupture. Thingbaijam and Mai (2016) demonstrated that a TEX distribution best characterizes the calculated distributions of earthquake slips in either an individual or an average sense. The authors examined the goodness of fit rigorously over TEX and other well-known distributions, including the EXP, Weibull, Gamma, and lognormal distributions, through the analysis of slip models of 190 earthquakes, and they concluded that the TEX distribution has the best fit to slips along fault planes, especially at larger slip values. Eq. (12) is exactly the same form as the TEX distribution suggested by Thingbaijam and Mai, where $u_c$ can be corresponded to $D/\gamma$ in Eq. (20) or $D/k\gamma$ in Eq. (21). However, the reasons for slip distributions being in TEX forms and rupture physics underlying the parameters of TEX remain unclear. Whereas Thingbaijam and Mai's work was a model-free analysis, here we aim to establish the stochastic dynamics for their empirical TEX distribution.

More insights can be obtained immediately from the proposed stochastic dynamics of earthquake ruptures. This paper presents a preliminary model based on the Langevin equation, focusing on a stationary state of rupture; however, the mechanisms that arrest a rupture are not considered. The dynamics themselves cannot predict when a rupture stops, and a rupture process is expected to deviate from the prediction of our model as it approaches the boundaries. According to the classical theory of tensile fracture for elastics, a growing Griffith-type rupture increasingly concentrates stress at the crack edges; therefore, an infinite perfectly elastic solid that fractures under homogeneous stress will not arrest until the boundaries are encountered (Kanninen and Popelar, 1985; Rundle et al., 1998). In the fault system, the frictional mechanisms near the boundaries must be different to arrest rupture. Thus, this model and its statistical features are valid only for sites that are away from boundaries. On the other hand, techniques of trimming the edges of the slip models are usually applied to remove superfluous low slips that may probably be artifacts of the earthquake rupture inversion (Somerville et al., 1999; Mai and Beroza, 2000). Especially in the study of Thingbaijam and Mai (2016), an extended data trimming process in fact improves the overall goodness of fit to the TEX distribution. Alternatively, according to the stochastic dynamics of earthquake ruptures proposed in this study, the TEX trend only reflects the steady-state part of the random process, which is relatively away from the boundaries. Therefore, the improvement of the goodness of fit may also be attributed to the unexpected removal of nonstationary slips near the boundaries.

Furthermore, Thingbaijam and Mai suggests a relationship between the average slip $u_{avg}$ and the fitting parameter $u_c$ of each model, summarized empirically as $log_{10} u_c = 1.05 log_{10} u_{avg} + 0.07$. Considering the resolution in rupture slip inversion, the exponent 1.05 is relatively close to unity. Thus, the link between $u_c$ and $u_{avg}$ can be established through stationary solutions [Eqs. (18) and (19)]. As demonstrated in Eqs. (22) and (23), the expected values of these two processes are $D/\gamma$ and $D/k\gamma$, respectively, which is the corresponding fitting parameter $u_c$ of TEX as shown in Eqs. (20) and (21).

$$P_{st}(v) = \frac{\gamma}{D} exp\left(-\frac{\gamma}{D}|v|\right), \langle v \rangle = \int_0^\infty v P_{st}(v) dv = \frac{D}{\gamma} \tag{22}$$

$$P_{st}(x) = \frac{\gamma k}{D} exp\left(-\frac{\gamma k}{D}|x|\right), \langle x \rangle = \int_0^\infty x P_{st}(x) dx = \frac{D}{k\gamma} \tag{23}$$

**Conclusions**

Various models describe earthquake ruptures as stochastic processes intrinsically involving all of the unknowns into random variables. The Brownian walk model in the research of Ide (2008) is a typical Langevin equation with viscous damping. Ide's Langevin equation describes the rupture process of slow earthquakes by defining the overall rupture area as the random variable. The model demonstrates realistic simulations of waveforms and moment rate functions by using reasonable values of model parameters. Another example is the traveling density wave model proposed by Rundle et al. (1996), which is a more complicated Langevin equation that illustrates earthquake events as sudden transitions toward minimal free energy of the fault system. The Langevin force in the equation of motion in their work represents the random disordered surfaces and other uncertainties of the fault system, and it produces irregular energy barriers, thus preventing sudden decay from definitely occurring in every cycle, leading to a more realistic model.

Developments in probabilistic tsunami or seismic hazard assessment have included the stochastic slip distributions of earthquakes to determine the overall probability of particular tsunami heights or ground-shaking levels (Geist and Parsons, 2006, 2009). The stochastic slip model quantifies variations in slip to reasonably estimate the probability of specified tsunami heights or ground-shaking intensities at individual locations resulting from a specific fault.

In this study, we presented a stochastic model based on the Langevin equation of frictions to investigate the underlying physics of earthquake ruptures. With a finite rupture duration, the original EXP distribution of slips along a fault plane can be truncated. Consequently, we demonstrated that the TEX distribution in the work of Thingbaijam and Mai can be physically attributed to the stationary solution of stochastic dynamics for the rupture process. The numerical simulation implies that the observation of a real earthquake rupturing can be considered a realization of microscopic-scale stochastic processes, accounting for the highly heterogeneous properties that compose earthquake faults, such as friction.

This study revealed that the fitting parameter $u_c$ in the TEX distribution of Thingbaijam and Mai can be related to the ratio of diffusion coefficient $D$ and friction coefficient $\gamma$ in the Langevin equation for earthquake rupture dynamics. However, a method must be formulated for interpreting $D$ and $\gamma$ in the fault system.

## Data and Resources

The reference rupture models were collected from SRCMOD, http://equake-rc.info/SRCMOD/. The effective source dimensions for the given slip model and the corresponding 2D rupture plane were computed using the study by Thingbaijam, K.K.S. (thingbaijam@gmail.com), http://www.equake-rc.info/cers-software/.

## Acknowledgments

This manuscript was edited by Wallace Academic Editing.